\documentclass[a4paper]{PoS}
\usepackage{bm}
\usepackage[flushleft]{threeparttable}

\title{New Advances in Pulsar Magnetosphere Modelling}

\ShortTitle{New Advances in Pulsar Magnetosphere Modelling}

\author{\speaker{Christo Venter}\\
        Centre for Space Research, North-West University, Potchefstroom Campus, Private Bag X6001, Potchefstroom 2520, South Africa\\
        E-mail: \email{Christo.Venter@nwu.ac.za}}

\abstract{The wealth of high-energy ($E \gtrsim 50$~MeV) and very-high-energy ($E > 100$ GeV) data accumulated over the past few years have provided unprecedented opportunities to probe pulsar emission models. The \textit{Fermi} Large Area Telescope (LAT) has now detected over 200 $\gamma$-ray pulsars of vastly different ages, providing an extensive dataset of spectra and light curves compared to a mere decade ago. Ground-based Cherenkov telescopes deepened our appreciation of the mysterious richness of the pulsar mechanism by a surprise detection of pulsed emission up to 1~TeV from the Crab pulsar. These discoveries position us to make real progress in pulsar theory. A number of studies have developed new and enhanced existing dissipative magnetohydrodynamical (MHD) and particle-in-cell (PIC) codes to solve the global electrodynamics and pursue fundamental questions about magnetospheric particle injection, acceleration, and radiation. While MHD models capture the global aspects of pulsar magnetospheres, the microphysics need to be probed self-consistently by PIC simulations. Some dissipative MHD models consider the current sheet (CS) as an important site for high-energy curvature radiation (CR), while early PIC results point to energy dissipation taking place in the CS, where particles are accelerated by magnetic reconnection and may possibly emit $\gamma$ rays via synchrotron radiation (SR). This is in marked contrast to some older local emission models that studied CR $\gamma$ rays produced in gaps nearer to the spinning neutron star. The universality of these results should become clearer as current computational restrictions are overcome and boundary conditions are refined. Continued and combined polarimetric, spectral, and temporal measurements should aid us in scrutinising these new emission models in our persistent pursuit of a deeper understanding of the pulsar marvel.}

\FullConference{4th Annual Conference on High Energy Astrophysics in Southern Africa\\
		25-27 August, 2016\\
		Cape Town, South Africa}

\begin{document}

\section{Introduction}
The year 2017 marks the golden anniversary of the discovery of (radio) pulsars \cite{Hewish68}. The pulsar journey has been exciting and rewarding, with many questions remaining after decades of research, but also much progress in the last years, both observationally and theoretically. The many unsolved mysteries will hopefully become clearer as more data are accumulated, computational power increases, and pulsar models continue to mature. By way of introduction, I summarise five basic questions in pulsar physics in Table~\ref{tab1}. There are many more, and for brevity many of the details have been suppressed. While some questions may have an obvious answer, the issue is that the details of such sweeping answers are usually lacking. I also list some ideas that might help us to make progress on these questions in the last column. Interestingly, uncertainties regarding the origin and acceleration of the magnetospheric plasma seem to be linked to many of these outstanding questions. In this review article, I will therefore focus my attention on some recent microscopic particle-in-cell (PIC) simulations in Section~\ref{sec:PIC} to highlight the progress that has been made in this area, in addition to providing a description of the latest dissipative magnetohydrodynamic (MHD) models (Section~\ref{sec:Dissipative}). 

There exist a number of complementary reviews, e.g., by Beskin~\cite{Beskin16} giving a chronological overview of the development of pulsar magnetosphere models, Caraveo~\cite{Caraveo14} providing an observational $\gamma$-ray pulsar overview, Melrose \& Yuen~\cite{Melrose16} as well as Cerutti \& Beloborodov~\cite{Cerutti16c} relating progress in the development of pulsar electrodynamics, Hirotani~\cite{Hirotani06} reviewing outer gap (OG) and slot gap (SG) models, Harding~\cite{Harding13} reviewing polar cap (PC) and SG models, Venter \& Harding~\cite{Venter14} assessing the impact of geometric light curve modelling on pulsar emission characterisation, Grenier \& Harding~\cite{Grenier15} focusing on the \textit{Fermi} LAT legacy plus implications for theory, Harding~\cite{Harding16} discussing the use of light curve modelling in vacuum, force-free (FF) and dissipative MHD magnetosphere models to probe pulsar magnetospheric structure as well as that of the emission region, Contopoulos~\cite{Contopoulos16} discussing the current sheet (CS), Y-point (Section~\ref{sec:FF}), and dissipation of energy, and Breed \textit{et al.}~\cite{Breed16} summarising the young field of very-high-energy pulsar science.

\begin{table}[!ht]
\hskip-1.2cm
\caption{Summary of some open questions in pulsar physics.}\label{tab1}
\vskip2mm
\begin{threeparttable}
 \centering
\begin{tabular}{|l|l|l|l|}
\hline
{\bf Question} & {\bf Answer} & {\bf Details} & {\bf Ideas}\\
\hline
\hline
1.\ What is the 	& Rotation 	& $\bullet$ Conversion of rotational energy 		& $\bullet$ Acceleration / radiation\\
\quad origin of	 	& (rotational 	& \quad to pulsed emission / winds? 			& $\bullet$ Spin-down\\
\quad pulsed		& energy)	& $\bullet$ What is the radiation mechanism?  		& $\bullet$ CR\tnote{1}, SR\tnote{2}, ICS\tnote{3},\,\,\,SSC\tnote{4}\\
\quad emission?		&		& $\bullet$ Which are the radiating particles? 		& $\bullet$ Electrons / positrons / ions\\
			&		& $\bullet$ How are these particles accelerated? 	& $\bullet$ $E$-fields; reconnection \\
			& 		& $\bullet$ How and where do they originate? 		& $\bullet$ PC charge \\
			& 		&  							& \quad extraction; pair cascades\\
\hline
2.\ Where does   	& Local models 	& $\bullet$ Gap formation. 				&  $\bullet$ Self-consistent and\\
\quad the radiation  	& (PC,OG,SG)	& $\bullet$ How are gaps sustained? 			& \quad sustainable gap formation\\
\quad come from? 	& or  		& $\bullet$ Dissipation in CS. 				& $\bullet$ Plasma microphysics \\
			& (return) CS	& 							& \quad (e.g., conductivity $\sigma^\prime$)\\
			&  		& 							& $\bullet$ Boundary conditions\\			
\hline
3.\ What is the 	& Dipolar near,	& $\bullet$ Corrections? Deviations? 			& $\bullet$ Sweepback / Multipoles?\\
\quad structure 	& toroidal	& $\bullet$ Local / global geometry?	 		& $\bullet$ Self-consistent global \\
\quad of the		& far		& $\bullet$ Spin-down?		  			& \quad models: FF vs.\ \\
\quad magneto-		& 	 	&  							& \quad dissipative\\
\quad sphere?		&  		& 					 		& $\bullet$ Uniqueness of solution? \\
\hline
4.\ Is there an 	& Indeed.  	& $\bullet$ Age or environment?				& $\bullet$ Suppression of pair\\
\quad evolutionary	& 		& $\bullet$ Details of recycling scenario?		& \quad formation with age\\
\quad sequence?		& 		& $\bullet$ Role of $B_{\rm LC}$\tnote{5}\,\,\,?	& $\bullet$ Different pulsar classes\\
			
\hline
5.\ What are the 	& Striped  	& $\bullet$ The `$\sigma$-problem':			&  $\bullet$ PWNe observations\\
\quad properties of	& wind, 	& \quad Conversion of EM\tnote{6}\,\,\,\,to		&  $\bullet$ MHD simulations\\
\quad the pulsar 	& particle- 	& \quad particle energy?				& \\
\quad wind?		& dominated	& $\bullet$ Particle leakage from PWNe\tnote{7}\,\,?	& \\
\quad  			& at large	& $\bullet$ Contribution to local leptonic		& \\
\quad  			& distances	& \quad cosmic-ray spectrum?				& \\
\hline
\end{tabular}

\begin{tablenotes}
            {\small 
            \item[1] Curvature radiation.
            \item[2] Synchrotron radiation.
            \item[3] Inverse Compton scattering.
            \item[4] Synchrotron-self-Compton emisson.
            \item[5] The magnetic field at the light cylinder $R_{\rm LC}$ where the corotation speed equals the speed of light in vacuum.
            \item[6] Electromagnetic.
            \item[7] Pulsar wind nebulae.}
        \end{tablenotes}
     \end{threeparttable}
\end{table}

\section{Modelling Assumptions: Plasma Density, Magnetospheric Structure, and Gaps}
Goldreich and Julian~\cite{GJ69} argued that the rotationally-induced $E$-field hugely dominates gravity (and particle inertia), so that a pulsar magnetosphere must to be filled with highly conductive plasma lifted from the stellar surface. This plasma is dense enough to screen out the accelerating component of the $E$-field parallel to the local $B$-field (i.e., $E_{||}$), so that the electromagnetic force on the plasma vanishes (i.e., FF conditions are applicable). However, this plasma is leaving the magnetosphere at relativistic speeds, so it must be replenished continuously, otherwise gaps with low local plasma density (and non-zero $E_{||}$) may form. Single-photon (magnetic)~\cite{Erber66,Sturrock71} and two-photon~\cite{Cheng86} pair production possibly hold the key to this replenishment of plasma. On the other hand, the screening can not be fully effective, since observations of pulsed high-energy emission require sites of efficient dissipation of energy and particle acceleration.

Historically, the questions regarding plasma density (i.e., is it uniform and is it dense enough to screen $E_{||}$?) have been approached in different ways. If one assumes that there is no plasma (e.g., when the binding energy in the stellar surface is very high), one is left with a neutron star surrounded by vacuum. Deutsch~\cite{Deutsch55} found the solution to this $B$-field called the rotating vacuum dipole (RVD) solution~\cite{Yadigaroglu97,Cheng00}, which has been used in several pulsar light curve models since~\cite{Dyks2004a,Venter09,Bai10a}. In this magnetosphere, there is no ambient charges or currents (and thus no hope for pulsed emission by accelerated charges). Other models also considered the (vacuum) static dipole solution~\cite{Harding78,Du10} or a GR\footnote{General-Relativistic}-corrected static dipole~\cite{MH97} $B$-field solution as the basic magnetospheric structure. On the other end of the spectrum, if there is dense plasma so that the $E$-field is screened everywhere (providing corotation of $B$-field lines with the star), one considers FF conditions and this scenario is characterised by the so-called `pulsar equation'. The latter has been solved for the aligned~\cite{Contopoulos99} and oblique~\cite{Spitkovsky2006} cases, also using full MHD ~\cite{Komissarov07,Tchekhovskoy13}. In this case, there can be no particle acceleration (Section~\ref{sec:FF}) and thus also no non-thermal pulsed emission. 

In reality, the conditions in a pulsar magnetosphere must be between the two extremes (vacuum vs.\ plasma-filled) mentioned above. Local gaps may form where there are deviations from FF conditions (i.e., the charge density deviates from the Goldreich-Julian one). Low-altitude PC models~\cite{Harding81,Daugherty96} (and its low-$B$-field extension, the pair-starved PC or PSPC model~\cite{HUM05}), OG models with gaps occurring above the null-charge surface (where the Goldreich-Julian charge density is zero) and close to the last open\footnote{These are $B$-field lines that touch the light cylinder at a cylindrical radius $R_{\rm LC}=c/\Omega$ (with $\Omega=2\pi/P$ the angular speed and $P$ the rotational period) and bound the closed-field-line region.} field lines~\cite{Cheng86,Romani95,Romani96}, as well as SG models~\cite{Arons83,MH03,MH04} with gaps that range from the stellar surface out to $R_{\rm LC}$ along the last open field lines (and also annular and core gap models~\cite{Du11}) have been studied (and reviewed) extensively. These gaps provide $E$-fields that accelerate particles locally, leading to pulsed high-energy emission. While such local models include plasma generation (injection) and acceleration, they are decoupled from the global magnetospheric structure, since they do not include the effect of plasma currents on the underlying $B$-field structure~\cite{Li12}. More recently, dissipative magnetosphere solutions~\cite{Li12,Kalapotharakos12b,Kalapotharakos14} (Section~\ref{sec:Dissipative}) have been obtained. In these solutions, there are thus charges, currents, and acceleration occurring in the pulsar magnetosphere. 

As an intermediate step, and also prompted by the high rate of new $\gamma$-ray pulsar discoveries by \textit{Fermi} LAT at the time, geometric light curve models such as the two-pole caustic\footnote{The term `caustic' in this context alludes to the concentration or bunching of photons at particular observer phases, forming bright features on the sky map of the projected radiation beam, and leading to bright peaks in the light curve when sampled by a particular observer's line of sight.} (TPC)~\cite{Dyks2003,Dyks2004a,Bai10a} and OG~\cite{Watters09,Venter09} geometries have been applied to constrain the pulsar emission sites as well as the pulsar geometry (i.e., its magnetic inclination angle $\alpha$ and observer angle $\zeta$ measured with respect to the spin axis), assuming the RVD field. Using the FF field as basis, OG, TPC, and separatrix layer (along the last open field line) models have also been studied~\cite{Bai10b}, although it seems that light curves produced by TPC and OG models in the RVD more closely reproduce the observed ones than light curves produced by FF magnetosphere models due to overprediction of the relative phase lag between radio and $\gamma$-ray light curves in the latter case~\cite{Harding11}. In such geometric models, a constant emissivity per unit length (in the corotating frame) is postulated along certain $B$-field lines, photons are assumed to be emitted tangentially to the local $B$-field lines, and aberration (when transforming from the corotating frame to the lab frame) plus time-of-flight delays (for photons emitted at different heights) are included. These models have taught us valuable lessons about preferred pulsar geometries and emission sites~\cite{Venter09,Venter12,Johnson14,Pierbattista15}, but also pointed to the fact that more work needs to be done to be able to describe the existing $\gamma$-ray pulsar light curves satisfactorily~\cite{Venter14}. 

While all these studies represented steady progress, the self-consistent arising of gaps without `local postulates' about the microphysics (e.g., local assumptions about plasma density or current distributions) is needed. PIC simulations are providing the first steps on this uncharted road (Section~\ref{sec:PIC}).

\section{Global FF Magnetospheres}
\label{sec:FF}
\subsection{The Static Axisymmetric FF Field}
A limiting and simplifying assumption for a pulsar's magnetospheric plasma is that it is conductive and dense everywhere, and that its energy density is small compared with the energy density of the EM field (plus the plasma drift velocity is subluminal, $E^2\leq B^2$, e.g.~\cite{Spitkovsky2006}), so that its properties may be described using FF electrodynamics (i.e., the effects of particle inertia may be ignored). Neglecting particle inertia and temperature (or pressure) yields
\begin{equation}
\rho\mathbf{E} + \frac{\mathbf{J}}{c} \times\mathbf{B}=0,
\end{equation}
with ${\rho}$ the charge density and $\bm{J}$ the current density. This constraint may be recast as the so-called pulsar equation~\cite{Michel73a,Scharlemann73,Okamoto74} or Grad-Shafranov equation, a non-linear partial differential equation that describes the flux of the poloidal $B$-field of an aligned rotator, with all other physical quantities describing the pulsar magnetosphere being related to the flux function $\Psi$, poloidal current $J$, and the angular speed $\Omega$~\cite{Timokhin06}. 
The above condition implies that the ideal MHD condition
\begin{equation}
\mathbf{E}\cdot\mathbf{B}=0
\end{equation}
is valid everywhere in the magnetosphere, i.e., corotation of $B$-field lines with the neutron star and $E_{||}=0$~\cite{Spitkovsky11}. 

There are a handful of analytic solutions, e.g., Michel~\cite{Michel73a} obtained the split-monopole solution to this equation (commonly believed to represent the magnetospheric structure farther from the star~\cite{Petrova16a}), as well as a solution for a corotating relativistic dipole magnetosphere with zero poloidal current valid inside $R_{\rm LC}$~\cite{Michel73b} (see also~\cite{Michel82}), while solutions also exist for a slightly perturbed monopole~\cite{Beskin98}, and also for an exact axisymmetric dipole with a differential rotational magnetospheric velocity distribution and general toroidal structure~\cite{Petrova16a,Petrova16b}. 

Contopoulos \textit{et al.}~\cite{Contopoulos99} found the first numeric solution to the pulsar equation for a dipole $B$-field near the star, allowing for a smooth transition of $B$-field lines across $R_{\rm LC}$ as well as for current closure. This solution is characterised by a corotating closed zone and a zone of open field lines with poloidal $B$-field lines becoming monopolar at large distances from the star where the toroidal $B$-field component dominates (Figure~\ref{fig:FF}a). Current flows out along the open field lines, and the major part of the return current flows in the CS and boundary / separatrix between the open and closed-field-line regions. At the tip of the closed zone, the CS splits into two at the so-called Y-point, which is in the equatorial plane and may be near $R_{\rm LC}$, but not necessarily~\cite{Goodwin04,Contopoulos05,Timokhin06,Spitkovsky11}. Gruzinov~\cite{Gruzinov05} studied this solution in more detail, including the energy losses of such an aligned rotator. 

Later time-dependent simulations~\cite{Spitkovsky2006,Komissarov06,McKinney06,Contopoulos10,Parfrey12} confirmed the existence of the stationary FF structure. However, the question of how the needed plasma is supplied to sustain the FF conditions (e.g., the detailed properties of the formation of secondary pair cascades~\cite{Timokhin06}) were not addressed.

\subsection{The Oblique FF Field}
\begin{figure}[t]
  \begin{center}
  \includegraphics[width=14cm]{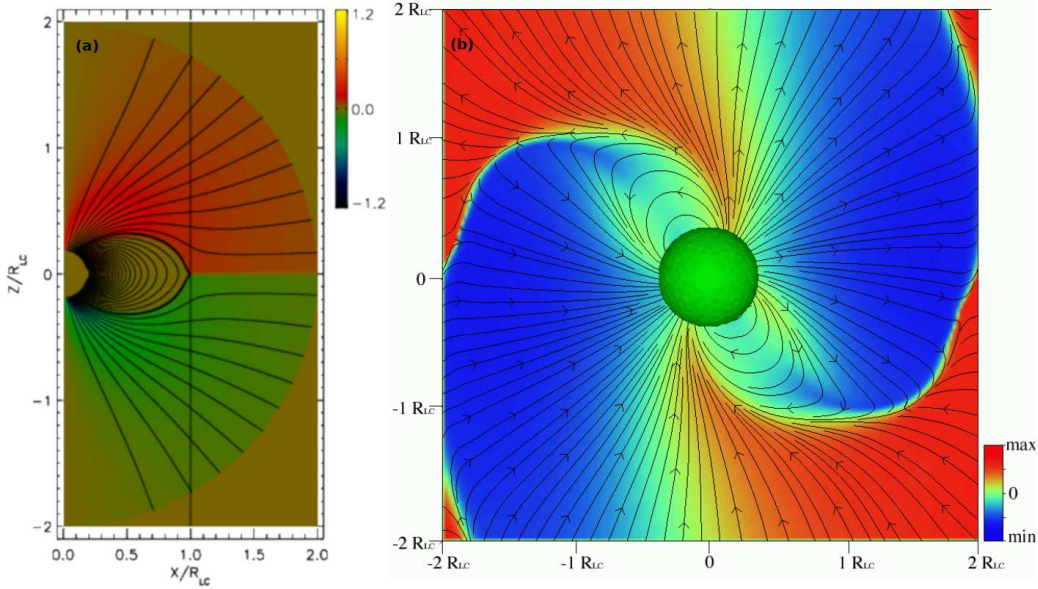}
  \caption{Snapshots of FF simulations of aligned (panel~a) and oblique (panel~b) rotators. In the first case, poloidal field lines of the steady-state solution are shown, with the thick black line indicating the last open field line touching $R_{\rm LC}$, and the colour indicating the normalised toroidal $B$-field component. In the second panel the magnetic inclination $\alpha=60^\circ$, and the $B$-field lines in the $\bm{\mu}-\bm{\Omega}$ plane in the corotating frame are indicated. The colour indicates the $B$-field perpendicular to the plane. From Spitkovsky~\cite{Spitkovsky2006}.}
  \label{fig:FF}
  \end{center}
\end{figure}
Spitkovsky~\cite{Spitkovsky2006} solved a system of Maxwell equations in flat spacetime satisfying the FF condition~\cite{Blandford02,Komissarov02} using a finite-difference time-domain method for both the axisymmetric and oblique rotator cases (similar results converging to the first numeric one~\cite{Contopoulos99} were independently obtained for both the aligned~\cite{Gruzinov05,McKinney06} and oblique~\cite{Kalapotharakos09,Contopoulos10,Kalapotharakos12a,Petri12a} cases). A slice of the fully 3D oblique-rotator solution is shown in Figure~\ref{fig:FF}b. In the $\bm{\mu}-\bm{\Omega}$ plane, with $\bm{\mu}$ the magnetic moment and $\bm{\Omega}$ the angular velocity vector, the $B$-field lines are similar to the Contopoulos \textit{et al.} solution~\cite{Contopoulos99}, but 
now there is an undulating CS (``ballerina skirt'') about the rotational equator (which was shown to be stable to distances of at least 10$R_{\rm LC}$ as well as for several stellar rotation periods~\cite{Kalapotharakos12a}), occurring in a wedge with opening angle $2\alpha$ and having a wavelength $2\pi R_{\rm LC}$. Similar to the inclined split-monopole solution~\cite{Bogovalov99}, the $B$-field lines in this plane become straight beyond $R_{\rm LC}$.
The Y-point still occurs near $R_{\rm LC}$. The generalised solution for the pulsar spin-down (integrated Poynting flux) was also obtained~\cite{Spitkovsky2006}:
\begin{equation}
  L_{\rm sd}\sim L_0\left(1+\sin^2\alpha\right),
\end{equation}
compared to the vacuum solution of $L_{\rm sd,vac}= (2/3)L_0\sin^2\alpha$, with $L_0=\mu^2\Omega^4/c^3$. 

\subsection{Striped-wind Emission Models and Properties of the CS}
\label{sec:striped}
The question regarding the physical properties of the CS is a vital one, especially since this region have been argued to plausibly be a prime location for the generation of high-energy pulsed emission~\cite{Contopoulos10,Philippov15a}. Indeed, Kalapotharakos \textit{et al.}~\cite{Kalapotharakos14} note that the physical conditions in the CS significantly modify the global magnetosphere structure. Typical questions are: Does magnetic reconnection\footnote{During reconnection, $B$-field lines change their topology to enter into a lower-energy configuration during which process magnetic energy is converted to kinetic energy through acceleration or heating of charged particles. Ideal MHD conditions are locally violated, and lines of force of opposite polarity can be broken and rejoined in a CS~\cite{Yamada10}.} take place there? Do plasmoids form (via the tearing instability)? Is there dissipation, and what is the nature thereof? What is the effect of internal thermal pressure on the CS thickness? Such questions have been pursued~\cite{Lyubarskii96,Petri11,Petri12b,Arka13,Uzdensky14,deVore14,Mochol15} in the context of so-called ``striped-wind'' models~\cite{Coroniti90,Michel94} (see P\'etri~\cite{Petri16} for a comprehensive review). 

\begin{figure}[t]
  \begin{center}
  \includegraphics[width=15cm]{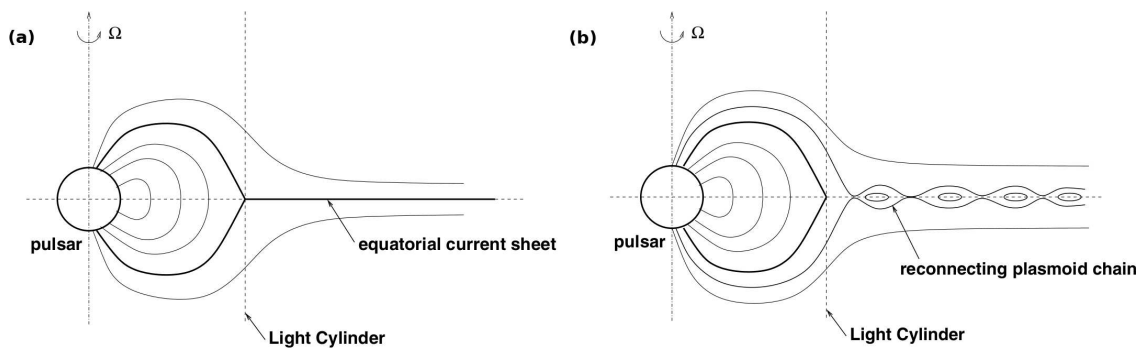}
  \caption{The basic axisymmetric magnetosphere (panel~a), and the tearing of the equatorial CS (formation of plasmoids, panel~b). From Uzensky \& Spitkovsky~\cite{Uzdensky14}.}
  \label{fig:Plasmoid}
  \end{center}
\end{figure}

Uzensky \& Spitkovsky~\cite{Uzdensky14} argued that the CS is a natural artefact of a rotating pulsar magnetosphere, and that magnetic reconnection dissipates a large fraction of the pulsar spin-down power there. Since the numerical FF codes cannot treat the CS properly, they developed a near-$R_{\rm LC}$ reconnection model to constrain the local plasma conditions. They argued that reconnection takes place violently via the formation of plasmoids (growing `magnetic islands'; Figure~\ref{fig:Plasmoid}b) of different sizes. These plasmoids are continuously forming and merging with one another and are ejected quasi-periodically. Magnetic energy dissipation takes place within small inter-plasmoid CSs. The relativistically hot reconnection layers occurring in pulsars are subject to strong SR  cooling, leading to significant plasma compression (density enhancement). They estimated that for the Crab pulsar, the temperature, pair plasma density, and layer thickness (comparable to the Larmor radius of relativistic electrons and positrons) are about 10~GeV, 10$^{13}$\,cm$^{-3}$, and 10~cm. For a relativistic Doppler factor of order 100, the first two of these quantities translate to typical lab-frame values of hundreds of GeV (so the particles can indeed radiate pulsed GeV emission by SR and pulsed TeV-band emission via inverse Compton scattering of ultraviolet or X-ray emission from the pulsar) and densities of order 10$^{15}$\,cm$^{-3}$. 

Physical models that constrain the properties of the CS, as described above, are complementary to the dissipative MHD (Section~\ref{sec:Dissipative}) and PIC (Section~\ref{sec:PIC}) simulations described below.

\section{Dissipative Magnetospheres}
\label{sec:Dissipative}
\subsection{Constant-$\sigma^\prime$ (Conductivity) Solutions}
It is clear that the FF limit is an idealistic one, since there can be no dissipation of energy and hence particle acceleration and high-energy emission in this framework. However, such magnetospheres do provide a global structure that is consistent with the stellar boundary condition of a perfect conductor, a smooth transition of $B$-field lines through $R_{\rm LC}$, and the formation of a large-scale MHD wind. They furthermore specify both the currents and the sign of charges flowing in the magnetosphere, but cannot provide any information on production of particles needed to sustain these currents~\cite{Kalapotharakos12b}. 

To more fully probe the locations where particle acceleration may take place, and the effect of such deviations from ideal MHD conditions on the magnetosphere structure, dissipative FF approaches were developed~\cite{Li12,Kalapotharakos12b} in which various prescriptions were invoked that led to non-zero accelerating $E$-fields, after which the structure of the magnetosphere was solved. For example, a finite conductivity $\sigma^\prime$ may be introduced, modifying the current density (although other prescriptions give qualitatively similar results)~\cite{Kalapotharakos12b}. This admits magnetosphere configuration solutions that cover the full spectrum from vacuum ($\sigma^\prime=0$) to FF ($\sigma^\prime\rightarrow\infty$). The CS survived the introduction of a non-zero $\sigma^\prime$, a non-zero $E_{||}$ developed over the PCs and along the separatrices (increasing with decreasing $\sigma^\prime$), and $B$-field lines closed beyond $R_{\rm LC}$. As $\sigma^\prime$ was increased, regions of large charge and current densities increased in size, the CS became more pronounced, $B$-field lines became straighter at large distances, and the amount of spin-down (Poynting flux) increased. 

Li \textit{et al.}~\cite{Li12} prescribed a simple form of Ohm's law (plus a choice of the fluid frame, i.e., a particular current density, relating it to the $E$-field and $B$-field), also introducing a constant $\sigma^\prime$ throughout the magnetosphere (this $\sigma^\prime$ is related to the maximum potential drop that a test particle can experience as it moves along the $B$-field lines). They found that the maximum out-of-plane $B$-field (with respect to the $\bm{\mu}-\bm{\Omega}$ plane) increased and the extent of the closed-field-line region shrank with increasing $\sigma^\prime$. A smooth, monotonic decrease in spin-down power (as a function of $\alpha$) was observed for increasing $\sigma^\prime$, due to an increase in the amount of sweepback and opening up of poloidal $B$-field lines. Intermittent pulsar phenomenology may potentially be described by identifying the observed ``on'' and ``off'' states with magnetospheres characterised by different $\sigma^\prime$ and hence different spin-down rates. Alternatively, such different states may reflect magnetospheres with different current densities (linked to pair formation processes)~\cite{Kalapotharakos12b}.

A next step is to include dissipative magnetosphere solutions in light curve models to connect the global $B$-field structure to observations. Using both a geometric plus emission approach (including CR reaction in the latter case), Kalapotharakos \textit{et al.}~\cite{Kalapotharakos12c} modelled the expected high-energy light curves of an orthogonal rotator as a function of $\sigma^\prime$. In the first approach, there was a broadening in peak widths (due to larger assumed gap widths) accompanied by a phase shift to the right (i.e., a lag) with increasing $\sigma^\prime$. These changes may be linked to changes in magnetospheric structure (i.e., increased sweepback for higher $\sigma^\prime$ producing PCs that are more offset from the magnetic pole). In the second approach, particle trajectories in the lab frame were approximately calculated, since the particles do not flow along $B$-field lines in the corotating frame as in the FF case. Generally, the broadest pulses corresponded to mid-$\sigma^\prime$ values (and narrowest pulses to highest $\sigma^\prime$), and a non-monotonic behaviour was observed for phase lags as a function of $\sigma^\prime$. At high $\sigma^\prime$ values, the distribution of $E_{||}$ was impacted more by a change in $\sigma^\prime$ than the magnetosphere structure was. Furthermore, at high $\sigma^\prime$, a significant part of the emission originated near the CS outside $R_{\rm LC}$.

Kalapotharakos \textit{et al.}~\cite{Kalapotharakos14} next constructed a macroscopic model based on dissipative magnetospheres, incorporating CR and a slightly new prescription for the current density (a new form of Ohm's law) plus different prescriptions for the $\sigma^\prime$ profile. They applied a very high $\sigma^\prime$ for the closed-field-line regions, and different constant values of $\sigma^\prime$ for the open $B$-field lines. They studied the actual trajectories of charged particles together with their acceleration by the self-consistently generated $E_{||}$ as well as their CR energy losses. For low $\sigma^\prime$, there was significant emission at lower altitudes inside $R_{\rm LC}$, while the radiation occurred at higher altitudes as $\sigma^\prime$ was increased (near the CS beyond $R_{\rm LC}$ where $E_{||}$ is higher, following the higher poloidal current density requirement there). The latter was due to the corresponding decrease in $E_{||}$ that implied longer acceleration distances for particles to acquire adequate energy to emit CR. Emission from the CS was additionally enhanced by relatively smaller values of radius of curvature for the associated particle trajectories that originate close to the PC rim. This emission was, however, not uniform near the CS. Small values of $\sigma^\prime$ were characterised by broad light curves (both single and double peaks, the latter occurring when $\alpha$ and $\zeta$ were large), while those corresponding to large $\sigma^\prime$ were narrower in some cases and morphologically more complex, since they contained emission from regions both inside and outside $R_{\rm LC}$. For very high $\sigma^\prime$, all emission originated beyond $R_{\rm LC}$ near the CS, and the peaks were narrow (and invisible for too low $\zeta$, similar to the standard OG model case). 

The main goal of Kalapotharakos \textit{et al.}~\cite{Kalapotharakos14}, however, was to reproduce the anti-correlation between the $\gamma$-ray peak separation $\Delta$ (i.e., phase difference between two main peaks) and the radio-to-$\gamma$-ray light curve phase lag $\delta$ (phase difference between the radio pulse and first $\gamma$-ray peak) demonstrated by the \textit{Fermi} LAT~\cite{2PC} for the population of known high-energy pulsars (and pointed out earlier as being a general property of outer-magnetospheric caustic light curve models~\cite{Romani95}). Assuming that the radio emission originated on the PC (i.e., at zero phase) and considering only younger pulsars, they derived $\Delta$ and $\delta$ for a number of $\alpha$ and $\zeta$ values. They found that the fits were not too good when assuming uniform distributions of $\alpha$ and $\zeta$ (later $F(\zeta)\propto\sin\zeta$ was also considered), but that the fits improved for higher $\sigma^\prime$, although they still could not reproduce the \textit{Fermi} observations satisfactorily. Moreover, in the latter case, the implied spectral cutoffs $E_{\rm cut}$ were too low to fit the data.

\subsection{The `FF-inside and Dissipative Outside' (FIDO) Model}
\begin{figure}[t]
  \begin{center}
  \includegraphics[width=13cm]{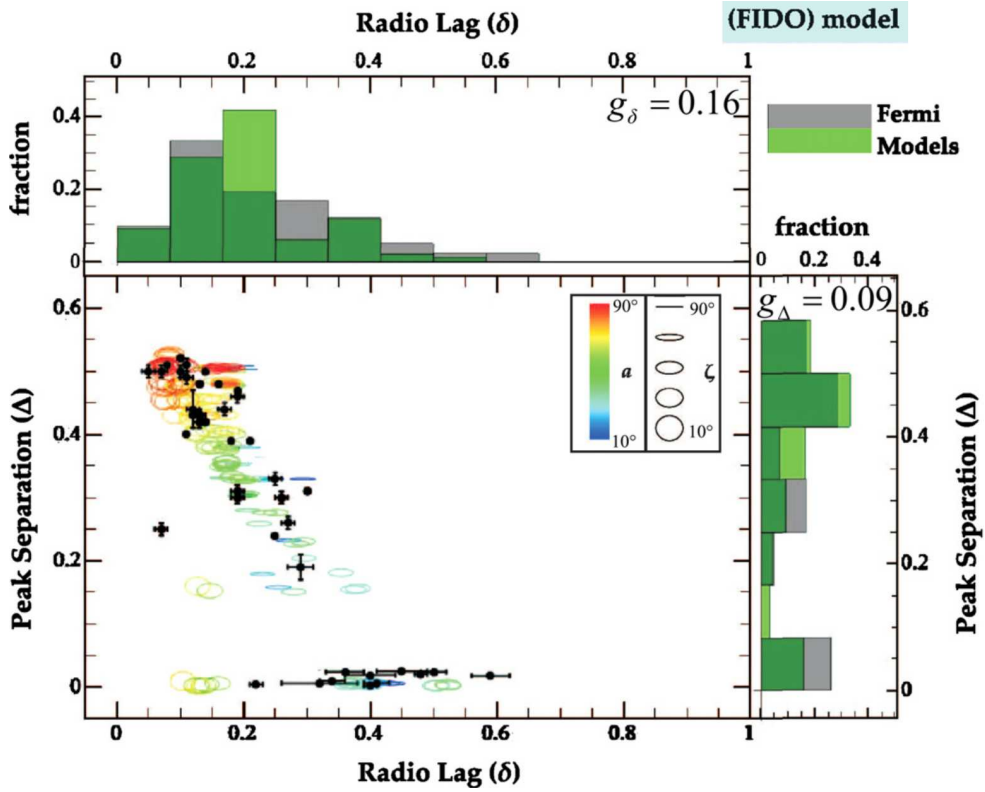}
  \caption{The $\Delta$-$\delta$ plot for younger pulsars in \textit{Fermi} LAT's Second Pulsar Catalog~\cite{2PC} together with results from the FIDO model. The colour indicates different $\alpha$ values, while the ellipses indicate different $\zeta$ values. In the histograms, grey indicates \textit{Fermi} data, while green indicates the FIDO results. From Kalapotharakos \textit{et al.}\cite{Kalapotharakos14}.}
  \label{fig:FIDO}
  \end{center}
\end{figure}
The fits of Kalapotharakos \textit{et al.}~\cite{Kalapotharakos14} to the $\Delta$-$\delta$ plot indicated that (spatially constant) low-$\sigma^\prime$ models were plagued by too large $\delta$ values. Assuming higher $\sigma^\prime$ values improved the situation, but did not solve it completely, leading to questions regarding the validity of assuming a constant $\sigma^\prime$ throughout the whole open-field-line region of the magnetosphere. Two solutions were attempted: one where a uniform emissivity per unit length in the CS was assumed for FF solutions (following particle trajectories from the rim of the PC out to $2R_{\rm LC}$), and one where $\sigma^\prime\rightarrow\infty$ inside $R_{\rm LC}$, but high (although finite) beyond $R_{\rm LC}$. 

Although the first attempt produced yet improved $\Delta$-$\delta$ fits, it still exhibited too large $\delta$ values. In the second case, dubbed the FIDO model, emission was produced near the CS (avoiding the emission component modulated by the inner-magnetosphere $E_{||}$), depending on the local conditions there. The FIDO model produced reasonable light curve shapes and furthermore provided very good fits to the $\Delta$-$\delta$ data (Figure~\ref{fig:FIDO}). The main reasons were that the emission was non-uniform around the CS (azimuthally as well as radially) such that emission was suppressed for small $\zeta$ (removing some higher-$\delta$ values from the plot), as well as being asymmetric across the CS, emphasising trajectories along the leading edge of the PC rim due to  such particles emitting at higher energies (again removing some higher-$\delta$ values). Next, a non-uniform distribution of $\alpha$ values could be inferred directly from the $\Delta$-$\delta$ fits using the FIDO model, pointing to a minimum probability at $\alpha\sim60^\circ$, and the maximum occurring at small $\alpha$ (consistent with alignment of the rotational and magnetic axes over time in such cases). Also, by choosing $\sigma^\prime\sim30\Omega$ (using units of angular speed $\Omega$ for $\sigma^\prime$) outside $R_{\rm LC}$, the $E_{||}$ was high enough so that the predicted spectral cutoffs occurred in the GeV band, as observed. The FIDO model could therefore successfully reproduce features of both high-energy light curves and spectra. 

Interestingly, the FIDO model may be viewed as containing elements~\cite{Venter14} of both the SG and OG models (i.e., it has a two-pole caustic geometry as in the first, and emission occurs exclusively at high altitudes as in the second, albeit at even higher altitudes than in the OG model) and its emission location is similar to that of other CS models (Section~\ref{sec:striped}), although its predicted emission pattern is quite different due to the non-uniform emissivity there. The suppression of $E_{||}$ in the inner magnetosphere furthermore implies that effective pair creation may occur along all open field lines (Section~\ref{sec:PIC}). 

\begin{figure}[t]
  \begin{center}
  \includegraphics[width=14cm]{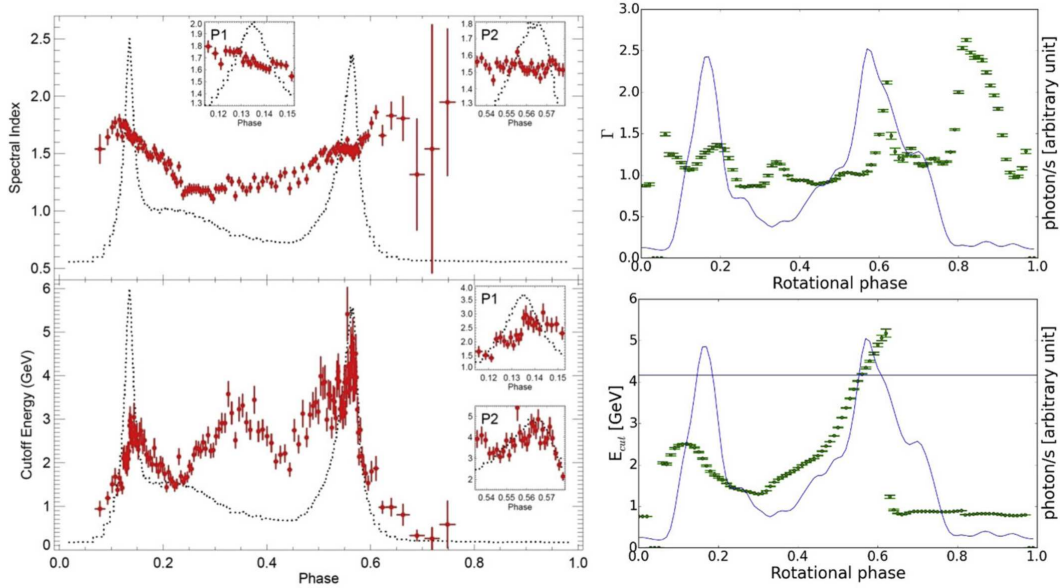}
  \caption{Observed (left) and predicted (right) light curves and phase-resolved spectral index $\Gamma$ (top) and cutoff energy $E_{\rm cut}$  (bottom) for the Vela pulsar. From Brambilla \textit{et al.}~\cite{Brambilla15}.}
  \label{fig:FIDO2}
  \end{center}
\end{figure}

A further test of the FIDO model was provided by Brambilla \textit{et al.}~\cite{Brambilla15} who modelled light curves, luminosities, phase-resolved spectra, and phase-averaged spectra (including flux and cutoff energy) due to CR for 1~260 different magnetosphere configurations (corresponding to different $\alpha$, $P$, surface $B$-fields, and $\sigma^\prime$ beyond $R_{\rm LC}$). These predictions were compared to the characteristics of eight bright \textit{Fermi} pulsars, yielding reasonable fits (e.g., Figure~\ref{fig:FIDO2}), although some features such as interpeak bridge emission or off-peak emission could not always be reproduced, and light curves satisfying all data constraints could not always be unambiguously found. In particular, they found that the best-fit $\sigma^\prime$ value for each pulsar increased with pulsar spin-down power and decreased with pulsar characteristic age. This is expected if more efficient pair production (i.e., implying a higher value of~$\sigma^\prime$) takes place in younger, more powerful pulsars, leading to more efficient screening of $E_{||}$ in this case. In general, the flux and spectral cutoff energy $E_{\rm cut}$ increased with $\zeta$ and decreased with $\sigma^\prime$ for a fixed $\alpha$. This is because larger $\zeta$ traces more closely the main caustic, while a larger $\sigma^\prime$ implies a lower $E_{||}$ and hence lower $E_{\rm cut}$. These quantities also increased with decreasing $P$ and increasing surface $B$-field, since $E_{||}$ increases in these cases. Increasing $\alpha$ resulted in a decrease of $E_{\rm cut}$ due to a lower $E_{||}$, as well as a decrease in flux due to a lower Goldreich-Julian charge density (with which the photon flux was scaled). 

There exists a general observational trend that $E_{\rm cut}$ is larger in phases corresponding to the second peak of a double-peaked pulse shape (e.g., for Vela~\cite{Abdo2010Vela} and the Crab~\cite{Aleksic2012}). This behaviour was reproduced in more than half of the configurations considered, implying that it may be common, but not universal. The trend of decreasing pulse width with energy~\cite{Abdo2010Vela,Aleksic2012} could, however, not be clearly reproduced. 

Despite failing to reproduce some details of the light curve phenomenology and phase-resolved spectral details, the FIDO model provided encouraging results in terms of capturing basic trends, and future refinements (e.g., considering a more realistic spatial distribution of $\sigma^\prime$ plus more configurations associated with more grid points in phase space) should improve its predictive power. Moreover, finding a model such as FIDO that is able to reproduce the $\gamma$-ray light curve phenomenology reasonably well impacts future microphysical simulations (Section~\ref{sec:PIC}) by providing a broad framework in which to fill in the details (e.g., origin of plasma to support the required densities and currents that lead to emitting regions plus emissivities that can successfully describe the data characteristics).

\section{PIC Simulations}
\label{sec:PIC}
Recently, a few groups have implemented \textit{ab initio} relativistic PIC simulations. As pointed out by Philippov \textit{et al.}~\cite{Philippov14}, the absence of a comprehensive theory for the radio and $\gamma$-ray emission from pulsars stems from a deficiency in our understanding of the plasma properties in various parts of the magnetosphere, including its origin (linked to the microphysics of pair production) and acceleration. Moreover, self-consistent modelling of plasma instabilities and particle acceleration (i.e., formation of `gaps' where FF conditions break down) is ultimately needed for a deeper understanding of the pulsar emission process, and this necessitates a kinetic treatment that will also allow the study of the CS as an integral part of the global magnetosphere. 

\subsection{The (Counter-)Aligned Rotator -- Charge-separated vs.\ Neutral Plasma}
\label{sec:P14}
The first self-consistent global simulation for an aligned/counter-aligned rotator was performed using a 2.5D PIC code~\cite{Chen14}. The star was spun up from $\Omega=0$ to a high angular velocity, and the evolution of the external EM fields and plasma was followed. Charge extraction from the stellar surface was invoked, coupled with acceleration of these primary particles and sparking of electron-positron discharge via pair conversion of CR photons emitted by the primary electrons. Two regimes were considered, one where pair conversion is efficient throughout the magnetosphere (e.g., through two-photon pair creation), and one where pairs are only formed close to the star (e.g., through one-photon pair formation). 

In the first case, a solution close to the FF one was obtained (e.g., exhibiting open/closed field line regions, asymptotically radial poloidal open field lines, the Y-point occurring close to $R_{\rm LC}$, and the formation of a CS). Roughly 10\% of current was carried by ions extracted from the neutron star at the footpoints of the CS. The discharge in the CS, however, relied on accelerated electron-positron pairs, requiring replenishment and thus leading to voltage oscillations (and ``breathing'' of the magnetosphere near $R_{\rm LC}$; see also Contopoulos~\cite{Contopoulos05}). This ``breathing'' was less pronounced for the counter-aligned rotator, since the CS extracted and accelerated electrons from the star instead of ions, aiding pair production so that the Y-point was nearly static. Far beyond $R_{\rm LC}$, magnetic reconnection was observed. This heated particles in an equatorial outflow and formed large plasmoids and kinks. Since pair creation above the PCs was not sustained (only weak acceleration of primary electrons occurred there for the aligned rotator), the boundary of the closed zone (converging in the Y-point) and especially the CS may thus be dominant locations for particle acceleration and generation of high-energy emission~(see also~\cite{Kalapotharakos14,Belyaev15}). 

For the second case, a burst of pair creation was observed when the star was spun up and a configuration similar to FF was attained only for a short time before $E_{||}$ that accelerated charges extracted from the surface (given the vacuum initial condition) was screened near the poles. Farther out, the magnetosphere resembled the well-known electrosphere or disc-dome configuration~\cite{Michel85b}. This solution will likely not converge to the Goldreich-Julian state, but rather become even closer to the electrosphere one. The above results indicate that pair formation in the outer magnetosphere is a necessary condition if the solution is to converge to an FF-like one. Thus, spin-down for pulsars with pair formation that exclusively takes place close to the star should depend on the spin and magnetic axes being misaligned.

An \textit{ab initio} relativistic 3D PIC code was applied to the case of a pulsar with aligned spin and magnetic axes~\cite{Philippov14}. Free escape of particles from the stellar surface was assumed and two extreme scenarios of plasma supply were considered (mimicking `dead' and `active' pulsars): (i)~particles were pulled from the stellar surface with no pair formation occurring in the magnetosphere; (ii)~neutral pair plasma was injected throughout magnetosphere (volume production) to model the effect of copious pair production that is thought to be the case in young pulsars. 

In the first scenario, a physical boundary condition on the stellar surface (assuming a constant interior $B$-field) was implemented to simulate unipolar induction, i.e., exterior $E$-fields were allowed to develop with components along the local $B$-field lines, and these $E$-fields extracted charges from the surface, thereby emulating space-charge-limited flow conditions. The solution reached the electrosphere configuration. In this charge-separated magnetosphere configuration, dome particles are trapped above the PCs and are corotating with the star. The disc particles are, however, not corotating and are unstable to the diocotron instability, which allows non-axisymmetric charge modulations and radial expansion of the disc so that the corotating Goldreich-Julian solution is approached within several rotation periods (although the FF magnetosphere is not fully achieved)~\cite{Petri02,Spitkovsky02}. During the simulation, only a small current flow occurred toward $R_{\rm LC}$, and the dipolar $B$-field structure was negligibly affected. Moreover, this flow did not cause the pulsar to spin down. 

In the second scenario, the initial magnetospheric configuration was devoid of plasma, but neutral pair plasma was injected at a fixed rate everywhere in space where the magnetisation $\sigma$ (ratio of Poynting to particle energy flux) was deemed large enough. A transition from the electrosphere solution to one very similar to the FF solution was observed. Less than 15\% of the Poynting flux was found to be dissipated within $2R_{\rm LC}$. (This is similar to independent findings~\cite{Belyaev15} that an FF structure is achieved when dense plasma is present throughout the magnetosphere. However, non-FF regions were found to exist where particles were accelerated, with the strongest dissipation occurring near the Y-point but also in the return current layers or RCLs and in the CS and even out to larger radii when $E_{||}$ was not fully screened in the outer magnetosphere. In this case a larger fraction of the spin-down luminosity was transferred to particles). No gap regions (with significant $E_{||}$) were observed near the poles, while particle acceleration could happen in the equatorial CS, possibly via magnetic reconnection. The drift-kink-instability was also observed in the CS. This study therefore reaffirmed the two limiting configurations of pulsars with suppressed vs.\ copious pair formation. It also highlighted the need for a refined treatment of pair production, as well as a generalisation to oblique rotators.

Interestingly, as an intermediate scenario to the ones described above, it was shown~\cite{Belyaev15} that even if an FF state is achieved by having dense plasma initially, such a configuration will collapse to the disc-dome state if pair production ceases in the magnetosphere, since the FF state cannot be sustained by injection of surface charge only, unless special circumstances exist (in terms of surface particle injection). On the other hand, Cerutti \textit{et al.}~\cite{Cerutti15} studied the transition from the disc-dome to FF configuration as a function of particle injection rate at the stellar surface, showing that if there is a high magnetisation at the stellar surface and particles are injected with a non-zero initial velocity, this may fill vacuum gaps with plasma leading to the FF solution. These studies therefore underscore the need for more realistic simulations of pair production to determine the detailed requirements for reaching different magnetospheric configurations.

\subsection{Oblique Rotator -- Improved Modelling of the Pair Formation Process}
In a follow-up work, PIC simulations of an oblique pulsar with improved modelling of electron-positron pair formation was presented~\cite{Philippov15a}. As noted previously~\cite{Beloborodov08,TA13}, particles in a space-charge-limited flow in regions where the current density along the $B$-field is below the Goldreich-Julian value (i.e., $J_{||}<J_{\rm GJ}=\rho_{\rm GJ}c$, with $\rho_{\rm GJ}$ the Goldreich-Julian charge density~\cite{GJ69}) do not attain high enough energies to produce electron-positron pairs. Thus, active pair creation can only occur where $J_{||}/J_{\rm GJ}<0$ or $J_{||}/J_{\rm GJ}>1$. In this case, bursts of pair production occur that produce high-density plasma with multiplicity factors (i.e., the ratio of spawned secondaries to primary charges) of up to $10^5$~\cite{TH15}. As before, particles (assumed to be positrons if they are positive, and electrons otherwise) were extracted from stellar surface. In conjunction with this injection of primary particles, two scenarios were considered: (i)~injection of neutral plasma everywhere in the magnetosphere; (ii)~a prescription approximating one-photon and two-photon pair production.

The first scenario is an extension of scenario~(ii) described in Section~\ref{sec:P14}, but for an oblique rotator. An FF-type solution was again obtained similar to an MHD one~\cite{Tchekhovskoy13}, with the open poloidal $B$-field lines becoming radial beyond $R_{\rm LC}$, the Y-point being located close to $R_{\rm LC}$, and with the CS oscillating about the equatorial plane. The drift-kink instability of the CS became negligible for magnetic inclinations of $\alpha\sim60^\circ$ and higher. The behaviour of the spin-down luminosity $L$ vs.\ $\alpha$ was also consistent with that found in MHD studies (i.e., $L\sim L_0(1+\sin^2\alpha)$ with $L_0\equiv\mu^2\Omega^2/c^3$).

In the second scenario, charges were injected from the full stellar surface. These primary charges were initially stationary but were accelerated as they moved into the magnetosphere. Whenever a primary particle's energy exceeded some minimal energy, an electron-positron pair was instantaneously formed (i.e., assuming zero mean free path and ignoring the effect of the local $B$-field strength and curvature). This was used as a proxy for regions of active pair formation. For the aligned rotator, the FF solution was approximated and the field became mostly toroidal beyond $R_{\rm LC}$, while the poloidal $B$-field lines become radial. Pair formation was suppressed above the PC region (where the current flow was sub-Goldreich-Julian), but did occur in the CS and RCLs. Strong $E$-fields in the CS led to the acceleration of parent particles there. A vacuum gap formed above the CS (see also~Chen \& Beloborodov~\cite{Chen14}). By confining pair creation to the inner magnetosphere or increasing the energy threshold above which pair creation occurs, the return current was suppressed and the disc-dome solution was recovered, confirming the results of Chen \& Beloborodov~\cite{Chen14}. 

For $\alpha \lesssim 40^\circ$, the solutions were similar to the aligned case: no pair formation took place above the PCs, where there was only weak acceleration and the bulk outflow was charge-separated. Pair formation only occurred in the CS and RCLs. However, for $\alpha > 40^\circ$, (quasi-periodic, partial) pair formation did occur in the PC zone, but also in the CS and RCLs. The $B$-field structure was similar to the FF solution. The CS beyond $R_{\rm LC}$ may thus be the dominant site for the production of high-energy photons (see also Cerutti \textit{et al.}~\cite{Cerutti15}). Nearly $\sim20\%$ of the Poynting flux was dissipated within $2R_{\rm LC}$ when $\alpha=0^\circ$, and this dropped to 3\% for $\alpha=90^\circ$, in agreement with earlier MHD results.

These results underscore the importance of considering high obliquities, but also raise the question of what would be the effects of considering curved spacetime, multipolar $B$-fields, ion extraction (especially for the case of orthogonal rotators), and increased spatial resolution.

\subsection{Aligned Rotator -- General Relativity (GR) \& Pair Formation}
\begin{figure}[t]
  \begin{center}
  \includegraphics[width=12.0cm]{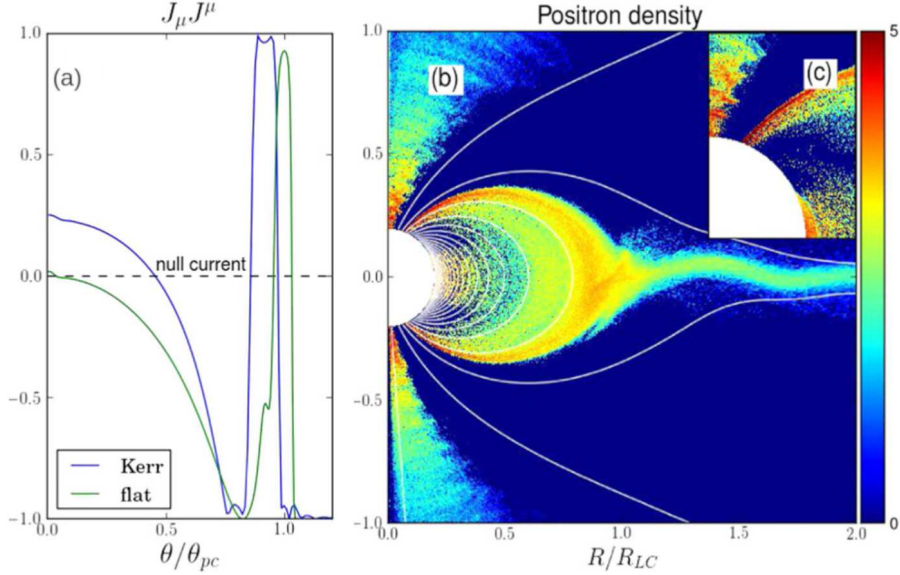}
  \caption{Panel~(a): Square of the four-current indicating its distribution over the PC, with $\theta_{\rm pc}$ the PC angle corresponding to the footpoint of the last open $B$-field line. This current becomes space-like near the rotational axis due to GR frame-dragging effects. Panel~(b) and (c): $\log$ of positron density divided by the Goldreich-Julian density at the pole, indicating non-stationary pair production taking place above the pole, as well as gaps that form near the CS. From Philippov \textit{et al.}~\cite{Philippov15b}.}
  \label{fig:GR}
  \end{center}
\end{figure}

Recent results~\cite{Philippov15a} indicated that low-obliquity pulsars only exhibit weak acceleration of primary particles near the poles and thus no pair production ensues in those sites due to the lack of high-energy radiated photons. The effect of GR corrections was next investigated~\cite{Philippov15b}, since it seems unlikely that such rotators would have suppressed pair formation -- ostensibly, pulsars with a range of inclinations are visible in radio and $\gamma$-ray wavebands (e.g., \cite{Johnson14,Pierbattista15}). An aligned rotator was studied incorporating a Kerr metric~\cite{Philippov15b}. The Goldreich-Julian charge density (which is set by local conditions) is decreased by the Lense-Thirring (frame-dragging) effect~\cite{Beskin90,MT92} compared to its flat-spacetime value, since the effective angular rotation speed is suppressed in the GR case. On the other hand, the time-averaged current density near the stellar poles are set by the twist of the $B$-field lines at $R_{\rm LC}$ (i.e., the global magnetospheric structure), and the latter is nearly unaffected by GR effects~\cite{Belyaev16}. These two facts lead to a relative increase in the ratio $J_{||}/J_{\rm GJ}>1$ in the GR case (making the current spacelike near the spin axis; see Figure~\ref{fig:GR}a), thereby restoring the pair cascade above the stellar poles.

Using an extension of a 2.5D axisymmetric PIC code to include GR effects, again injecting neutral plasma from the stellar surface, and modelling pair formation to occur once an energy threshold was exceeded, Philippov \textit{et al.} \cite{Philippov15b} found that for stars with a compactness above a critical value of $2GM/c^2R>0.5$ (with $M$ the stellar mass, $G$ the gravitational constant, and $R$ the stellar radius), non-stationary pair production occurred above the PC but also in the RCLs and equatorial CS (Figure~\ref{fig:GR}, panels b and c). A gap-like structure formed near the CS. The critical value of compactness necessary for pair formation to be triggered may be lowered if an additional source of plasma exists in the gaps that form near the CS. The oscillatory pair production cycles produced oscillations in the local EM fields, which may be connected to radio emission. The drift-kink instability was lastly suppressed when radiative cooling was taken into account, although it was still present. 

The fact that GR effects are essential for pair production in the aligned rotator was confirmed independently by Belyaev \& Parfrey~\cite{Belyaev16}. They used an analytical model that incorporates a fit to the distribution of poloidal $B$-field lines from FF studies~\cite{Tchekhovskoy16} coupled with the condition for one-photon pair production (i.e., $J_{||}/J_{\rm GJ}<0$ or $J_{||}/j_{\rm GJ}>1$) to study the location of pair-producing $B$-field lines above the PC in an FF magnetosphere (also see~\cite{Gralla16}). They noted that pair formation only occurred over a fraction of the PC closer to the pole, even with GR effects included. This conclusion also holds for weaker assumptions about the $B$-field topology that departs from the strict dipole assumption (e.g., for PCs displaced from the rotation axis, assuming that spatial variations occur on a scale much larger than the size of the PC). However, two-photon pair production may still occur in regions where pairs would otherwise not form (via the one-photon pair production mechanism) and may fill any outer gaps that may form with plasma.

\subsection{Light Curves, Spectra, and Polarisation in PIC Models}
Using 3D PIC simulations, Cerutti \textit{et al.}~\cite{Cerutti16a} self-consistently modelled the oblique magnetospheric structure, particle acceleration, and emission. They injected dense plasma (at twice the local Goldreich-Julian density in the form of pairs with an initial poloidal velocity of $0.5c$) from the rotating stellar surface, so that it fills the entire magnetosphere, leading to a quasi-FF configuration. About $10^8$ macro-particles filled the magnetosphere and emitted `macro-photons' via CR and SR. In these high-multiplicity solutions, it emerged that synchro-curvature emission from the PC regions led to MeV-band radiation, forming broad single-peaked light curves (since no significant particle acceleration occurred there due to efficient screening of $E_{||}$). Conversely, SR that peaks in the GeV band originated beyond $R_{\rm LC}$ at the Y-point and CS, being emitted by quasi-monoenergetic particles there.\footnote{However, for low-multiplicity solutions acceleration may take place in gaps that may form, since the formation and shape of such emission regions should be a function of the particle injection assumptions~\cite{Belyaev15}.} Thus, most of the dissipation took place within a few $R_{\rm LC}$, where relativistic magnetic reconnection converted magnetic energy to particle kinetic energy. 

\begin{figure}[t]
  \begin{center}
  \includegraphics[width=13.5cm]{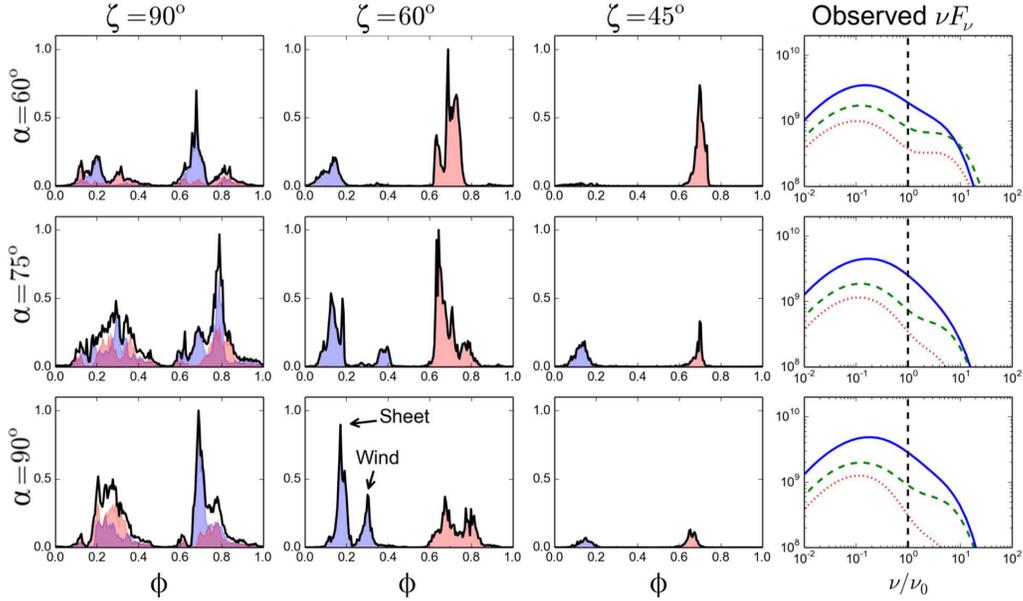}
  \caption{Three left columns: A sample of high-energy light curves for observer angles $\zeta=90^\circ,60^\circ,45^\circ$ and magnetic inclination angles $\alpha=90^\circ, 75^\circ, 60^\circ$. Filled blue lines indicate emission by positrons, filled red lines emission from electrons, and black solid lines the total emission from both species. Right-most panel: phase-averaged spectra from both electrons and positrons, with blue solid lines indicating $\alpha=90^\circ$, green dashed lines $\alpha=60^\circ$, and red dotted lines $\alpha=45^\circ$. From Cerutti \textit{et al.}~\cite{Cerutti16a}.}
  \label{fig:LC}
  \end{center}
\end{figure}

Since accelerated electrons and positrons flowed in opposite directions in the CS (positions flowing outward and electrons returning to the star), they were responsible for distinct emission patterns. Positron emission occurred over the full CS leading to the formation of a caustic, and double-peaked light curves were formed when the observer's line of sight traversed this caustic. Two additional secondary peaks originated from the wind region in high-obliquity pulsars, where energetic positrons emitted radiation beyond $R_{\rm LC}$. On the other hand, electron emission was mostly confined near the Y-point, where they cooled rapidly due to the relatively strong perpendicular $B$-field there. Thus, at intermediate $\zeta$, the first peak was mainly due to positron emission, while the second was due to electron emission (see Figure~\ref{fig:LC}). Single-peaked light curves occurred at lower $\zeta$ and intermediate $\alpha$, due to electron emission in the CS. Both particle and emitted photon spectra were strongly dependent on $\alpha$ and $\zeta$. 
Sub-pulse variability due to plasma waves and tearing and kink instabilities in the CS were also observed, but may not be observable in practice. Future simulations including full pair production may well exhibit high-energy emission from the RCLs. The total radiation efficiency (for all $\zeta$) decreased with increasing $\alpha$, dropping from 9\% to 1\% as $\alpha$ increased from $0^\circ$ to $90^\circ$; however, these numbers may increase for particular $\zeta$ values due to beaming.

Polarisation may potentially be a valuable discriminator between models when used in conjunction with the predicted spectral and temporal features, since different radiation mechanisms coupled to different locales of emission may leave unique detectable imprints on the observed polarisation properties of pulsars. While it has become standard to use the rotating vector model (RVM)~\cite{Radhakrishnan69} and more modern implementations of this model\footnote{See, e.g., the work of P\'etri~\cite{Petri16b} for an analytic treatment of the phase-resolved polarisation angle for a general off-centred dipole configuration, leading to the so-called `decentred RVM' or DRVM. He also showed that this polarisation angle additionally depends on the emission altitude, contrary to the RVM case. Lyutikov~\cite{Lyutikov16} furthermore introduced another relativistic modification to the RVM, that of rotation of the direction of polarisation by a moving source, in addition to aberration and time-of-flight effects.} in the study of radio pulses and to constrain pulsar geometry (i.e., $\alpha$ and $\zeta - \alpha$), high-energy polarisation is a promising but relatively new avenue of research~\cite{Slowikowska09,Gros16}. 

Cerutti \textit{et al.}~\cite{Cerutti16b} modelled the phase-resolved polarisation properties of pulsed high-energy SR originating in the CS, directly calculating the Stokes parameters using their 3D PIC code~\cite{Cerutti16a}. They found that the emission was mildly polarised and that there was a clear anti-correlation between flux and degree of linear polarisation (the average degree of linear polarisation is $\sim15\%$ at on-pulse and $\sim30\%$ at off-pulse peak intervals), the latter effect being more evident at lower $\alpha$. Such depolarisation in peaks is reckoned as a signature of caustic emission, as has already been pointed out by Dyks \textit{et al.}~\cite{Dyks2004a}. Photons originating at different positions in the magnetosphere with different local $B$-field orientations (e.g., different relative strengths of the poloidal vs.\ toroidal $B$-field components or due to small-scale turbulence) arrive simultaneously at the observer so that the polarisation is suppressed, akin to destructive interference. Also, similar to Dyks \textit{et al.}~\cite{Dyks2004a}, rapid swings in the polarisation angle was observed near the peaks, linked in this case to the change of $B$-field orientation (``polarity flip'') as the observer's line of sight passed through the CS or ``striped wind'' region~\cite{Petri05}. At pulse maximum, both Stokes parameters $U<0$ and $Q<0$. Future X-ray and $\gamma$-ray polarimetry missions may therefore play an important role to discriminate between competing pulsar emission models by probing potentially unique polarisation features~\cite{Harding17}.

\section{Conclusions}
There has been a flurry of activity in the last several years pertaining to pulsar physics. Much of the recent development in pulsar magnetosphere models has been driven by high-energy and very-high-energy observations, although we are indebted to observers in all wavebands for their valuable contributions. 

Much progress has also been made on the theory front. The FF MHD models capture the global structure of the magnetosphere and therefore provide more realistic constraints on aspects of the emission geometry of the pulsar. The current and charge density are vital quantities that should be matched to microphysical conditions that are thought to exist (and will become clearer in future) in the magnetosphere. While the Poynting flux and hence the pulsar spin-down can be determined, the magnetosphere is filled with plasma and hence there is no acceleration, emission, or dissipation in these models. The question about the uniqueness of such solutions is also a concern. 

More realistic magnetospheres, including the FIDO model, take energy dissipation into account. These models indicate that CR from the CS may be important for production of high-energy emission. However, these models need further development, and in particular prescriptions for the microphysics such as the conductivity $\sigma^\prime$ at different positions in the magnetosphere. PIC simulations may therefore be a new and complementary tool to study and constrain the microphysical conditions in addition to the global ones, i.e., allowing for a self-consistent treatment of pair formation, particle acceleration, and radiation from first principles (influenced by the choice of boundary conditions). 

Early PIC results are pointing to the CS as an important dissipative region, which is in marked contrast to several older models that focused on the open field zone (or PC zone), although it was argued~\cite{Belyaev16} that a thin gap might still survive above the RCLs. Furthermore, magnetic reconnection may hold the key to particle acceleration, and the favoured emission mechanism may even be SR rather than CR (what has been the standard assumption for a long time), although this deserves further investigation. However, pulsar PIC simulations are still in their infancy~\cite{Beskin16}, and computational restrictions force us to use low spatial resolution (e.g., a small number of cells or large ratio of stellar to light cylinder radius), short runtimes, low particle numbers, low magnetisations, low multiplicities, low Lorentz factors, low ion masses, and low $B$-field line curvature radii. The effect of these restrictions and the robustness of results as the restrictions are eventually lifted remain to be seen. 

Apart from computational restrictions that hamper our access to the full (PIC) parameter space, another level of approximation is the neglect or simplification of the intricate details of certain radiation processes and the feedback thereof on the $B$-fields and $E$-fields in the pulsar magnetosphere. One example is the neglect of Quantum Electrodynamical (QED) effects that profoundly impact continuum radiation processes such as SR and IC as well as physical processes such as one-photon pair production~\cite{DH83}, vacuum polarisation, and photon splitting~\cite{HB97} for very high $B$-fields~\cite{Harding03}. 

In view of above retrictions, one is cautioned~\cite{Petri16} that there is still a lot of ground to cover to reconcile the microphysics and the global structure of plasma magnetospheres, so that rather than being a means to its own end, PIC simulations should encourage more critical and novel thinking. In a similar vein, Contopoulos~\cite{Contopoulos16} suggests that at this early stage, future PIC simulations should continue to study the CS and Y-point while combining them with ideal MHD or FF electrodynamic simulations of the global magnetosphere. All theoretical tools at our disposal should therefore be used to paint a fuller picture of the pulsar phenomenon. Such a hybrid / comprehensive approach is also needed in observational science: maximum model constraining power will be obtained by pursuing and combining ever more detailed polarimetric, spectral, and temporal measurements, guiding our way as we continue to pursue the grand problem of pulsar emission.

\acknowledgments
I acknowledge fruitful discussions with Alice Harding and Zorawar Wadiasingh. This work is based on the research supported wholly by the National Research Foundation (NRF) of South Africa (Grant Numbers 87613, 90822, 92860, 93278, and 99072). The Grantholder acknowledges that opinions, findings and conclusions or recommendations expressed in any publication generated by the NRF supported research are those of the author, and that the NRF accepts no liability whatsoever in this regard.

\end{document}